\documentclass[fleqn,twoside]{article}
\usepackage{espcrc2}

\usepackage{graphicx}

\newcommand{\AmS}{{\protect\the\textfont2
  A\kern-.1667em\lower.5ex\hbox{M}\kern-.125emS}}

\def\lapproxeq{\lower .7ex\hbox{$\;\stackrel{\textstyle <}{\sim}\;$}}
\def\gapproxeq{\lower .7ex\hbox{$\;\stackrel{\textstyle >}{\sim}\;$}}

\title{Extensive Air Showers and Accelerator Data -- The NEEDS Workshop}

\author{Ralph Engel\address{Bartol Research Institute,
        Univ. of Delaware, Newark, DE 19716, USA.}%
        \thanks{Highlight talk given at the Int. Symposium on Very High
                Energy Cosmic Ray Interactions, Geneva, Switzerland, 
                July 15 - 20, 2002.}%
        \thanks{present address: Forschungszentrum Karlsruhe,
        Institut f\"ur Kernphysik, Postfach 3640, 76021 Karlsruhe, Germany.}
}

\begin{document}

\begin{abstract}
Very high energy cosmic rays are typically studied by measuring
extensive air showers formed by secondary particles produced in
collisions with air nuclei. The indirect character of the measurement
makes the physics interpretation of cosmic ray data strongly dependent
on simulations of multiparticle production in showers. In April 2002
about 50 physicists met in Karlsruhe to discuss various aspects of
hadronic multiparticle production with the aim of intensifying the
interaction between high energy and cosmic ray groups. 
Current and upcoming possibilities at accelerators for measuring features of 
hadronic interactions of relevance to air showers were the focus of the
workshop. This article is a review of the discussions and conclusions.
\end{abstract}

\maketitle


\section{Introduction}

The interpretation of most cosmic ray
experiments relies on particle physics measurements done at accelerators.
This is obvious in the case of measurements of cosmic rays with energies
higher than $10^{14}\,$eV. At such energies direct measurements are very
difficult or impossible because of low statistics due to the rapidly
decreasing primary cosmic ray flux and limited detector aperture
\cite{Seo02a}.
However, utilizing the Earth's atmosphere as target, 
large detector apertures and
observation times can be achieved, extending the reach in energy up to
$10^{20}\,$eV and beyond.
The drawback is the highly indirect method of
measurement which is based on the detection of secondary particles,
forming extensive air showers (EAS), and associated
Cherenkov or fluorescence light. 

The complexity of EAS requires the
detailed simulation of hadronic and electromagnetic particle cascades.
Whereas there is a good understanding of the predictions of QED on 
em. particle production, up to now, hadronic multiparticle production cannot be
calculated on theoretical grounds. Although QCD is the accepted theory
of strong interactions, only processes with large momentum transfer
(hard scale) can be calculated reliably in perturbation theory. 
The majority of hadronic interactions
are not characterized by a hard scale and do not fall into
the domain of perturbative QCD. Therefore soft hadron production has to
be simulated using phenomenological models. Naturally, due to the lack
of a calculable theory, the predictions of currently used 
models on multiparticle production differ considerably. 
Measurements of 
hadronic interactions at fixed target and collider experiments are
the ultimate and most efficient method to learn more about soft particle
production. They are essential for tuning hadronic interaction models and
reducing their uncertainties when extrapolated to ultra-high energy.

\begin{figure}[!htb]
\includegraphics[width=7.5cm]{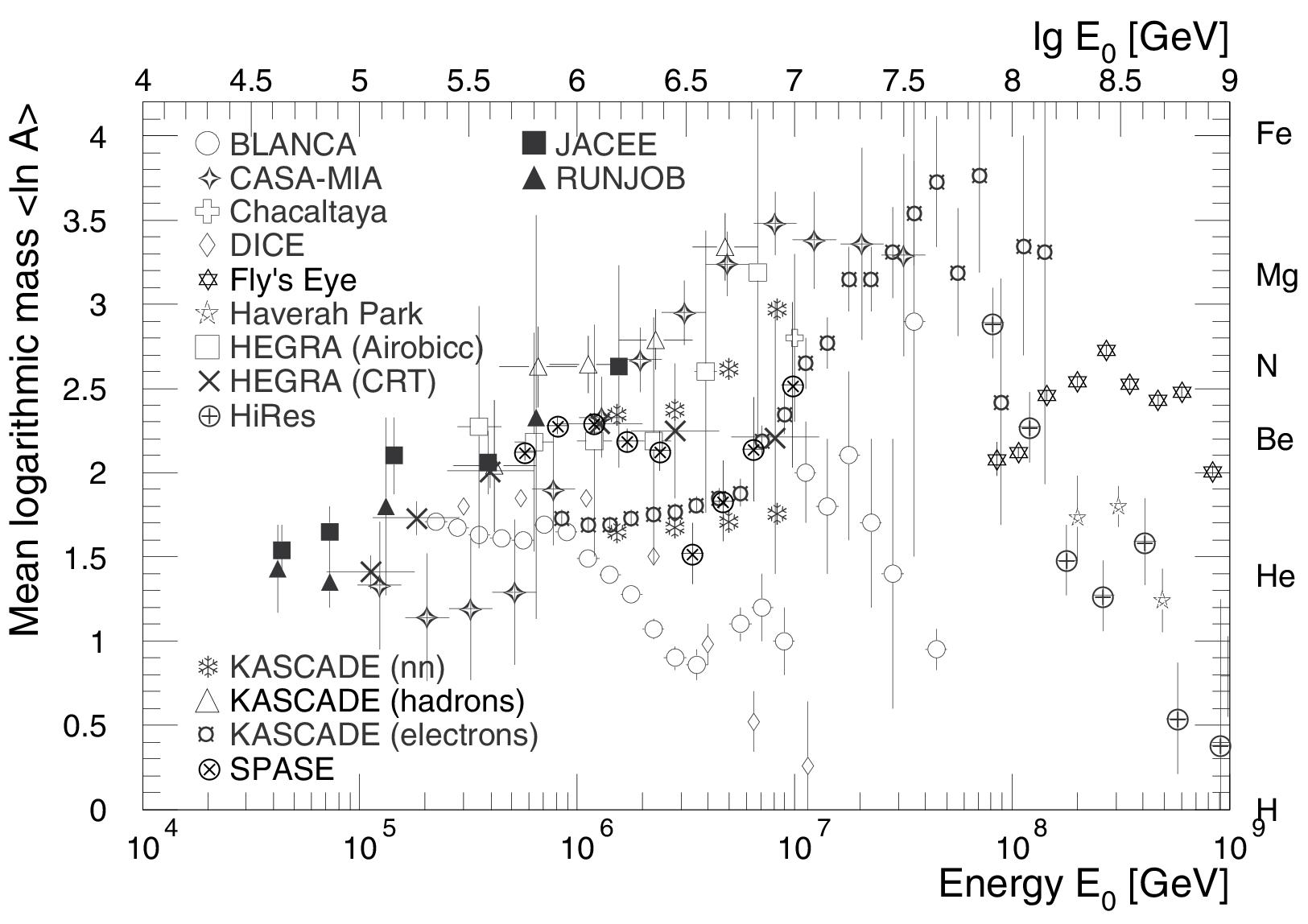}
\vspace*{-1.2cm}
\caption{Mean logarithmic mass of cosmic rays 
(from \protect\cite{Hoerandel01a}, modified). The data are derived from
published EAS measurements using the QGSjet interaction model
\protect\cite{QGSjet98}. For details see 
\protect\cite{Hoerandel01a,Hoerandel02a}.
}
\vspace*{-5mm}
\label{fig:lnA}
\end{figure}
There are many open questions related to the primary cosmic ray
spectrum. For example, the sources of the cosmic rays, the
origin of the knee at $3\times 10^{15}\,$eV
and the ankle at about $3\times 10^{18}\,$eV, to name but a few, are not
known.
Many EAS measurements have been
made to determine the mass composition
of the cosmic rays in this energy region, which would help understand
the
origin of the knee.
A compilation of the results expressed as mean logarithmic mass
$\langle \ln A\rangle$ is shown in Fig.~\ref{fig:lnA}.
To date there is no consistent picture of the
compositional changes in the knee energy region, however there is
general, qualitative agreement that the composition becomes heavier 
above $E\sim
3\times 10^{15}\,$eV. One of the main reasons for the discrepancies
between the different results is the use of different hadronic
interaction models in the analysis of the measurements (see also 
discussions in \cite{Swordy02a,Engel01a,Hoerandel02a}).

There is urgency of improving the interaction between the 
high energy physics (HEP) and cosmic ray (CR) communities.
Modern CR experiments have reached the statistics and 
precision that the
simulation of hadronic interactions becomes one of the 
main limiting factors of the data analysis
(for example, \cite{Roth02a,Gaisser02a}). 
Even small differences in
the assumptions on hadronic particle production in forward direction
are of crucial importance for the analysis. On the other hand, many
measurements of forward particle production, being most important for
the simulation of particle cascades, can be done at current
accelerators of moderate energy. In the course of concentrating all 
HEP capacities to few very high
energy collider projects more and more low and medium energy experiments
cease operation and will no longer be available for such measurements.

Discussing how accelerator measurements can help in understanding
CR data, about 50 physicists met in Karlsruhe for the workshop
{\it Needs from Accelerator Experiments for the Understanding of
High-Energy Extensive Air-Showers} (NEEDS). The workshop was organized by
Hans Bl\"umer, Andreas Haungs, Heinigerd Rebel (Forschungszentrum
Karlsruhe), and Lawrence Jones (Univ. of Michigan) and took place in
the Research Center Karlsruhe on April 18 - 20, 2002. 
The following list gives an overview of the contributions.
\begin{itemize}
\item
current situation regarding EAS and CR measurements\\
{\small
(M. Roth, J. Knapp, E.C. Loh, O. Saaveedra, G. Schatz,
A. Haungs, M. Risse, M. Unger)
}
\item
relation of hadronic interactions to EAS observables and hadronic
interaction models\\
{\small
(L. Jones, D. Heck, T. Stanev, R. Engel, S. Ostapchenko, J.N.
Capdeville, G. Battistoni, J. Ranft)
}
\item
relevant data from current accelerator experiments\\
{\small
(CDF @ Tevatron: V. Tano; H1 \& ZEUS @ HERA: A. Rostovtsev, M. Erdmann;
BRAHMS, PHOBOS, STAR \& PHENIX @ RHIC: D. Bucher, E941 @ AGS: B.
Fadem, HARP @ CERN PS: K. Zuber, G. Barr)
}
\item
planned accelerator experiments\\
{\small
(TOTEM/CMS \& ATLAS @ LHC: S. Tapprogge; CASTOR @ LHC: A. Angelis;
MIPP @ Tevatron: C. Rosenfeld)
}
\end{itemize}
Further details, including presentations of the speakers,
can be found on the workshop web page \cite{NEEDS-web}. 

\begin{figure}[!htb]
\includegraphics[width=7.5cm]{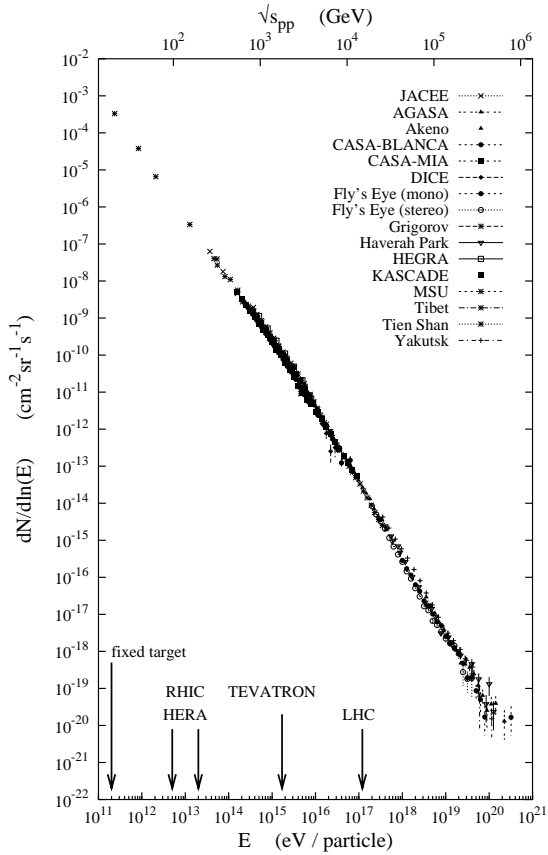}
\vspace*{-1.2cm}
\caption{Cosmic ray all-particle flux. The equivalent center-of-mass
energy is calculated for primary protons.
}
\vspace*{-5mm}
\label{fig:flux}
\end{figure}

This article is an attempt to review the discussions of this
workshop. 
In Sec.~\ref{sec:energies} a brief overview of the relevant
interaction energies and regions of phase space of secondary particles
is given. The current situation and problems of measuring primary CR
energy and mass spectra are reviewed in Sec.~\ref{sec:CR-part},
including the possibilities to assess hadronic interaction models. A
number of accelerator experiments and measurements of relevance to EAS
physics are discussed in
Sec.~\ref{sec:HEP-part}. Some conclusions are presented in the last part
of this article.

A summary of similar activities prior to this workshop can be found
in \cite{Jones01a,Jones01b}.


\section{Simulation of CR interactions: energies and phase space regions
\label{sec:energies}}

Fig.~\ref{fig:flux} shows measurements of 
the primary CR flux and equivalent collider energies. For RHIC and
LHC only the proton-proton collider option is shown. 
A detailed list of different acceleration
options and their equivalent CR energies can be found in
\cite{Jones01b}. The energy of cosmic rays spans an energy range of
more than 10 orders of magnitude and exceeds by
far that currently available at man-made accelerators.

\begin{figure}[!htb]
\includegraphics[width=7.5cm]{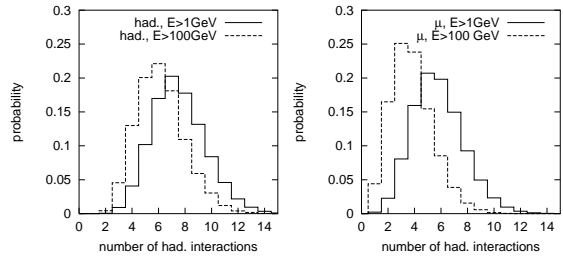}
\vspace*{-1.2cm}
\caption{
Number of subsequent interactions (generations) that give rise to
hadrons and muons of different energy at sea level. 
}
\vspace*{-5mm}
\label{fig:gen}
\end{figure}
Interactions of highest energy cosmic rays open a window to ultra-high
energy particle physics. In principle the analysis of EAS can yield 
information on multiparticle production in p-air collisions with CMS
energies of up to 400 TeV. However, the features of the first interaction of
a primary particle are largely washed out by the successive interactions
of the secondaries with correspondingly lower energies.
For example, Fig.~\ref{fig:gen} shows the number of successive
interactions (generations) in a proton induced shower of
$E=10^{15}\,$eV, which lead to the muons and hadrons observed at
sea level. To obtain sensitivity to the physics of the first few
interactions of the primary particle a good understanding of all
subsequent interactions is needed.

\begin{figure}[!thb]
\centerline{\includegraphics[width=7cm]{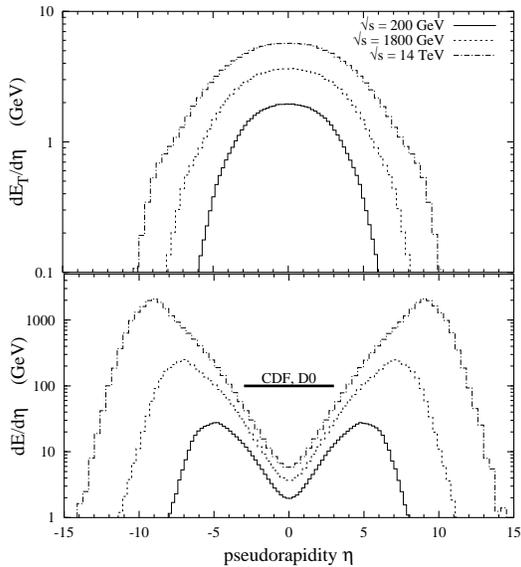}}
\vspace*{-1cm}
\caption{
Energy and transverse energy densities in proton-proton collisions as
function of pseudorapidity. The main acceptance region of the
Tevatron detectors CDF and D\protect$\emptyset$ (Run I) 
is indicated by the horizontal line.
}
\vspace*{-5mm}
\label{fig:eflow}
\end{figure}
One would expect that, because the energy of the knee corresponds
roughly to
that of the Tevatron collider, there is little uncertainty in modeling
hadronic interactions up to this energy. This is not the case as
modern collider experiments are designed to measure quantities 
which can be predicted within perturbation theory. 
Mainly
hadronic processes with at least one hard scale (large mass or high
virtuality) are studied. The measurement of hard processes 
requires typically high beam luminosities and sensitivity to secondaries
with large transverse momentum. By contrast, particle production in 
cosmic ray cascades is dominated by the most energetic particles with
small transverse momenta. The situation is shown in
Fig.~\ref{fig:eflow} by comparing
the energy flow in p-$\rm \bar p$ collisions at different collision energies
with the phase space covered by the CDF and D$\emptyset$ detectors.
In addition, measuring particles with momenta close to the beam direction is
technically challenging. The mean transverse momentum of the secondary
particles rises only very slowly with collision energy. This means that 
the detectors have to be placed the closer to the beam pipe the higher
the energy of the collision is to cover a
similar phase space region.

In summary, EAS and more generally cosmic ray interactions can only 
be understood and analyzed
successfully if hadronic interaction models are developed, tuned
to data, and maintained that\\
(i) cover the entire range of relevant energies extending 
from the particle production 
threshold to the energy of the primary cosmic ray,\\
(ii) give a good description of particle production in the forward
direction, i.e. soft and diffractive interactions, and\\
(iii) allow the extrapolation of accelerator measurements to higher
energies and to unmeasured phase space regions.


\section{Cosmic ray data and interaction models
\label{sec:CR-part}}

At the workshop numerous analyses demonstrated the strong dependence of the
interpretation of cosmic ray data on the assumptions on hadronic multiparticle
production. In the following we will concentrate on the
impact of the hadronic interaction model on the composition analysis in the
knee energy region. 

\begin{figure}[!htb]
\centerline{\includegraphics[width=6cm]{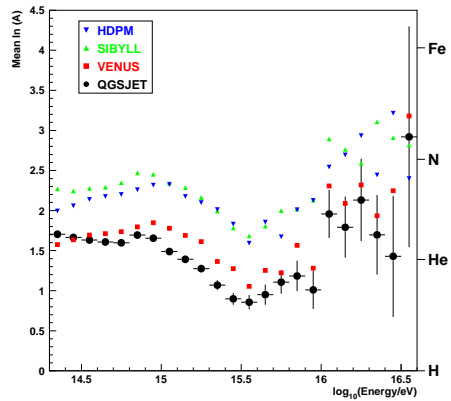}}
\vspace*{-1cm}
\caption{
Mean logarithmic mass as determined from CASA-BLANCA data using different
interaction models \protect\cite{Fowler00a}.
}
\vspace*{-5mm}
\label{fig:comp}
\end{figure}
The mass composition derived by the CASA-BLANCA Collab. \cite{Fowler00a}, 
shown in
Fig.~\ref{fig:comp} as mean logarithmic mass, is a typical example of
the model-dependence of EAS results. Using different hadronic
interaction models for analyzing the same EAS data leads to
significantly different conclusions on the mass composition. 
The statistical uncertainties are much smaller than the systematic
uncertainty due to the model dependence. It is clear that some of the
used hadronic interaction models describe collider data better than
others and a critical evaluation of the models is needed. First steps
towards a systematic comparison of models were done in \cite{Knapp96a}.

However, shortcomings of the modeling of hadronic multiparticle
production are obvious even if the QGSjet model is used, 
which provides currently the best
description of the data of the multi-component detector KASCADE.
A similar spread of $\langle \ln A\rangle$
values is found if different observables of the same
data set are analyzed with this model. For example, Roth et al. find a range of 
$\langle \ln A\rangle \approx 1.7 - 3$ for $E \sim 10^{16}\,$eV, whereas
an analysis based on electron and muon numbers gives a systematically
lighter composition than an analysis using muon and hadron observables
\cite{Antoni02a}.

\begin{figure}[!htb]
\includegraphics[width=7.5cm]{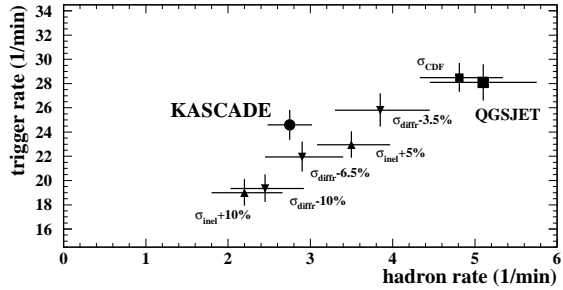}
\vspace*{-1.2cm}
\caption{Mean trigger and hadron rates of the KASCADE detector. Shown are
the observed rates and predictions obtained by varying the inelastic
cross-section in QGSjet. The point labeled by \protect$\sigma_{\rm CDF}$
corresponds to a \protect$\rm p$-\protect$\rm \bar p$ total 
cross-section as measured by CDF \cite{Tevatron-cs}. The other points
are labeled by the relative change of the cross-section as compared to the
QGSjet default extrapolation.
}
\vspace*{-5mm}
\label{fig:rates}
\end{figure}
On the other hand, the quality of the description of CR data by
simulations can also be used to characterize how well a given
interaction model describes leading particle production. For example, 
the KASCADE
detector allows the simultaneous measurement of many different
quantities which can be used as model discriminator
\cite{Antoni99a,Antoni01a}.
In particular the correlation of hadrons to other observables is
sensitive to the assumptions made in the interaction models. 
The inelastic proton-air cross-section is a quantity that strongly
influences the rate of hadrons observed in the detector. Increasing
the cross-section not only reduces the number of hadrons reaching sea
level but also reduces the predicted KASCADE trigger rate, which also
depends on the number of muons, see Fig.~\ref{fig:rates}.

Another
example is the analysis of Pamir emulsion chamber data
in \cite{Haungs01a}. Due to the high
altitude of the detector (4370m, $X_{\rm det}\approx 600\,$g/cm$^2$) 
earlier stages of the shower evolution and also
lower primary energies can be measured.
\begin{figure}[!htb]
\centerline{\includegraphics[width=6.5cm]{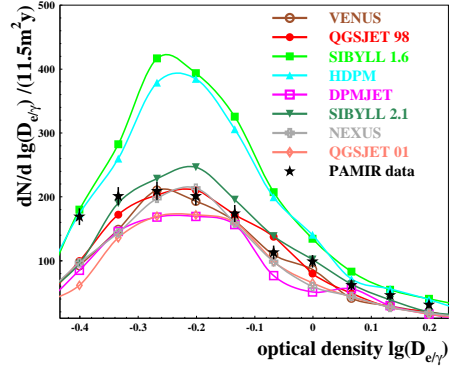}}
\vspace*{-1cm}
\caption{
Distribution of optical density calculated for the Pamir emulsion
chamber \protect\cite{Haungs01a}. Simulations for different
hadronic interaction models are compared to Pamir data.
}
\vspace*{-5mm}
\label{fig:pamir}
\end{figure}
Fig.~\ref{fig:pamir} shows the measured optical density distribution
of high-energy secondary shower particles 
observed in the Pamir emulsion chamber experiment 
together with
model predictions. Where available in CORSIKA \cite{Heck98a}, 
old and new model versions are shown. 
Larger optical density corresponds to higher
particle energy. The energy threshold is about 4 TeV for
photons and electrons, and 8 TeV for hadrons.

\begin{figure}[!htb]
\centerline{\includegraphics[width=7cm]{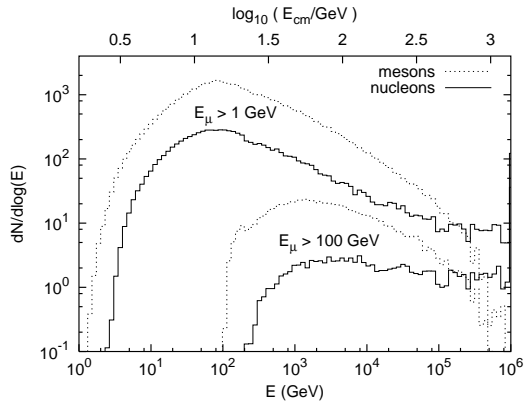}}
\vspace*{-1.2cm}
\caption{
Energy of ``last interaction´´ producing muons that reach sea
level \protect\cite{Engel00b}. Shown are the spectra of vertical EAS,
induced by protons with \protect$E=10^{15}$ eV, for muons with energies greater
than 1 and 100 GeV.
}
\vspace*{-5mm}
\label{fig:moth-en}
\end{figure}

Finally it should be mentioned that the simulation of hadronic
interactions at low energy is an important, integral part of any air shower
simulation. In particular the number of GeV muons, one of the important
energy estimators in EAS experiments, depends directly on the
secondary particle multiplicity in $\pi$- and p-air 
interactions in the 100 GeV range.
The typical energies probed in EAS muon production are given in
Fig.~\ref{fig:moth-en}. The histograms show the distribution of the energy 
of hadron $h_1$
inducing the ``last''
interaction producing a hadron $h_2$ that subsequently decays into a
muon which reaches
sea level ($X=1033$ g/cm$^2$)
\begin{equation}
h_1 + {\rm air} \rightarrow h_2+X;\hspace*{1cm} h_2 \rightarrow \mu +
X^\prime .
\end{equation}
Most of the muons are produced in collisions with energies about
10 to 100 times larger than the muon energy. In addition, more than 80\%
of the muons are produced in pion-air and not p-air collisions. A
recent analysis of interaction characteristics can be found in
\cite{Drescher02a} and muon measurements are compared to simulations,
for example, in \cite{Haungs02a}.


\section{Present and future accelerator experiments
\label{sec:HEP-part}}

\begin{figure}[!htb]
\centerline{\includegraphics[width=6.5cm]{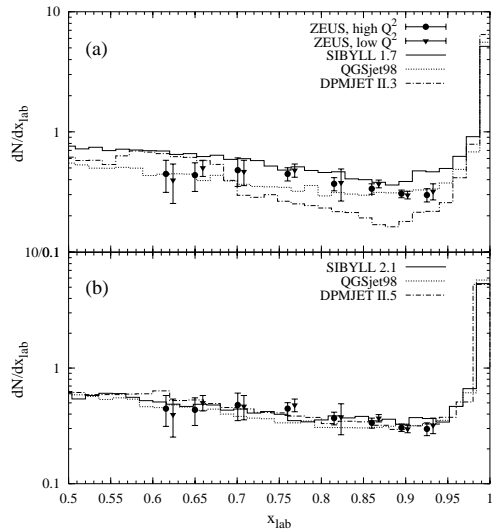}}
\vspace*{-1cm}
\caption{
Leading proton distributions in p-p collisions at
\protect$\sqrt{s}=200\,$GeV. The ZEUS data \protect\cite{Solano00a} 
refer to $\gamma^\star$-p
interactions measured in two different photon virtuality ranges and
\protect$\sqrt{s}_{\gamma^\star p} = 150 - 200\,$GeV (\protect$E \approx
2\times 10^4$ GeV). The upper panel (a) shows predictions of models
before the data became available \protect\cite{Engel98a} and the lower panel (b)
shows the results of the tuned models.
}
\vspace*{-5mm}
\label{fig:lead-pro}
\end{figure}

Parton density measurements at the HERA collider are of particular 
importance as input in
all modern hadronic interaction models. They are the basis of the
calculation of inclusive minijet cross-sections at high energy and are
one important component for the multiplicity and cross-section
extrapolations (see, for example, \cite{Engel01b}). In addition,
recently measured leading proton and neutron spectra are of direct
relevance to EAS simulations. Assuming that the forward, leading baryon
distributions in p-$\gamma$ collisions are independent of the target
type one can compare HERA data to model predictions for p-p
collisions. The HERA data are the first measurement of leading baryons
at energies greater than 400 GeV (Fig.~\ref{fig:lead-pro}). 

Tevatron measurements at $\sqrt{s}=1800\,$GeV are a benchmark for all
models. Unfortunately particle distributions and multiplicities
measured so far are restricted to the pseudorapidity 
range shown in Fig.~\ref{fig:eflow}. The measurement of hadron
distributions (p, $\pi$, and K) at large Feynman $x$ would allow a
direct model comparison. The large theoretical uncertainty in the model
extrapolations could be reduced by a measurement of rates and
inclusive cross-sections of jets with a transverse energy as low as
$5\,$GeV. Indeed it is the poorly known minijet cross-section which is one
of the major parameters in contemporary models. First steps in this
direction are the analysis of the soft underlying event in collisions
with high-$p_\perp$ jets \cite{Tano02a} and the measurement of events
with multiple jets \cite{Abe97b,Abazov02a}. The CDF and D$\emptyset$
detector upgrades for Run II include the installation of forward
detector components \cite{Alves02a}. The measurement of inclusive hadron
distributions, not only that of diffractive events, would be of great 
help for tuning EAS interaction models. It would also allow to reduce
uncertainties due to other phenomenological assumptions made in
simulation codes \cite{Ostapchenko02a}.

RHIC data on Au-Au collisions at $200\,$GeV/n have underlined the
limited predictive power of modern simulation programs. 
The observed central particle densities were about 20 - 30\%
lower than the theoretical expectations. RHIC heavy ion data are of
interest to cosmic ray simulations as they offer a cross check of
the theoretical concepts implemented in the simulation codes.
Furthermore the high parton densities in heavy nuclei at RHIC energy 
are expected to be comparable to that in light nuclei at correspondingly
higher energy. Heavy ion experiments typically select events according to the
centrality of the collision whereas for EAS simulations minimum bias
measurements are preferred. Some of the detectors have coverage of a
part of the forward direction \cite{RHIC}. 
The BRAHMS and STAR detectors allow particle
identification up to $\eta \approx 3.7$. Multiplicities can be measured
with PHOBOS up to 5.5 in pseudorapidity. 
A particularly interesting option would be the installation of
N$_2$ or O$_2$ gas targets.  

The new RHIC data have raised a
number of questions and competing model approaches are developed for 
their explanation. A discussion of recent RHIC results can be found in
 \cite{Klein-ISVHECRI} and their importance
for EAS simulations is analyzed in \cite{Ranft-ISVHECRI}.

\begin{figure}[!htb]
\centerline{\includegraphics[width=7cm]{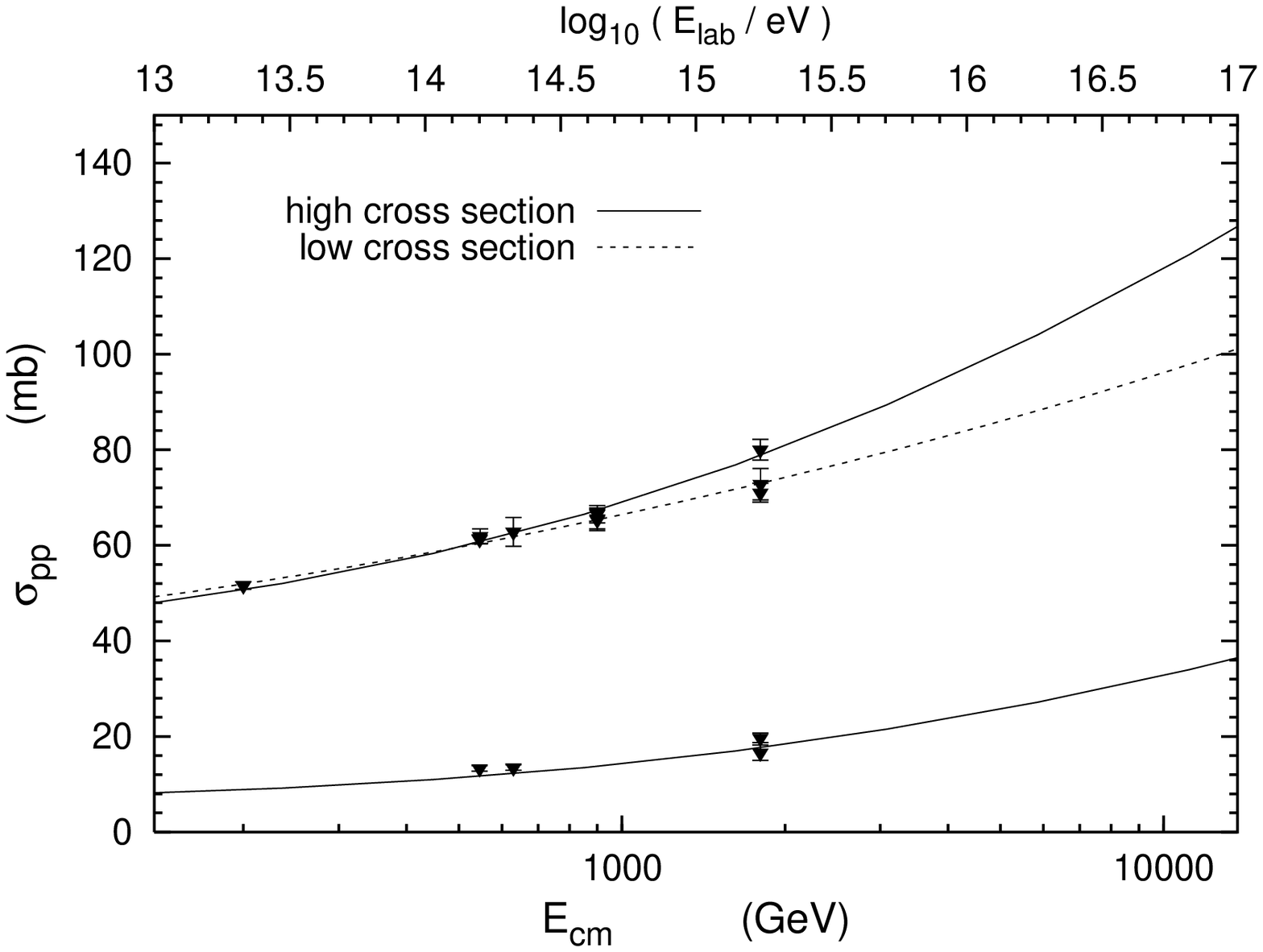}}
\centerline{\includegraphics[width=7cm]{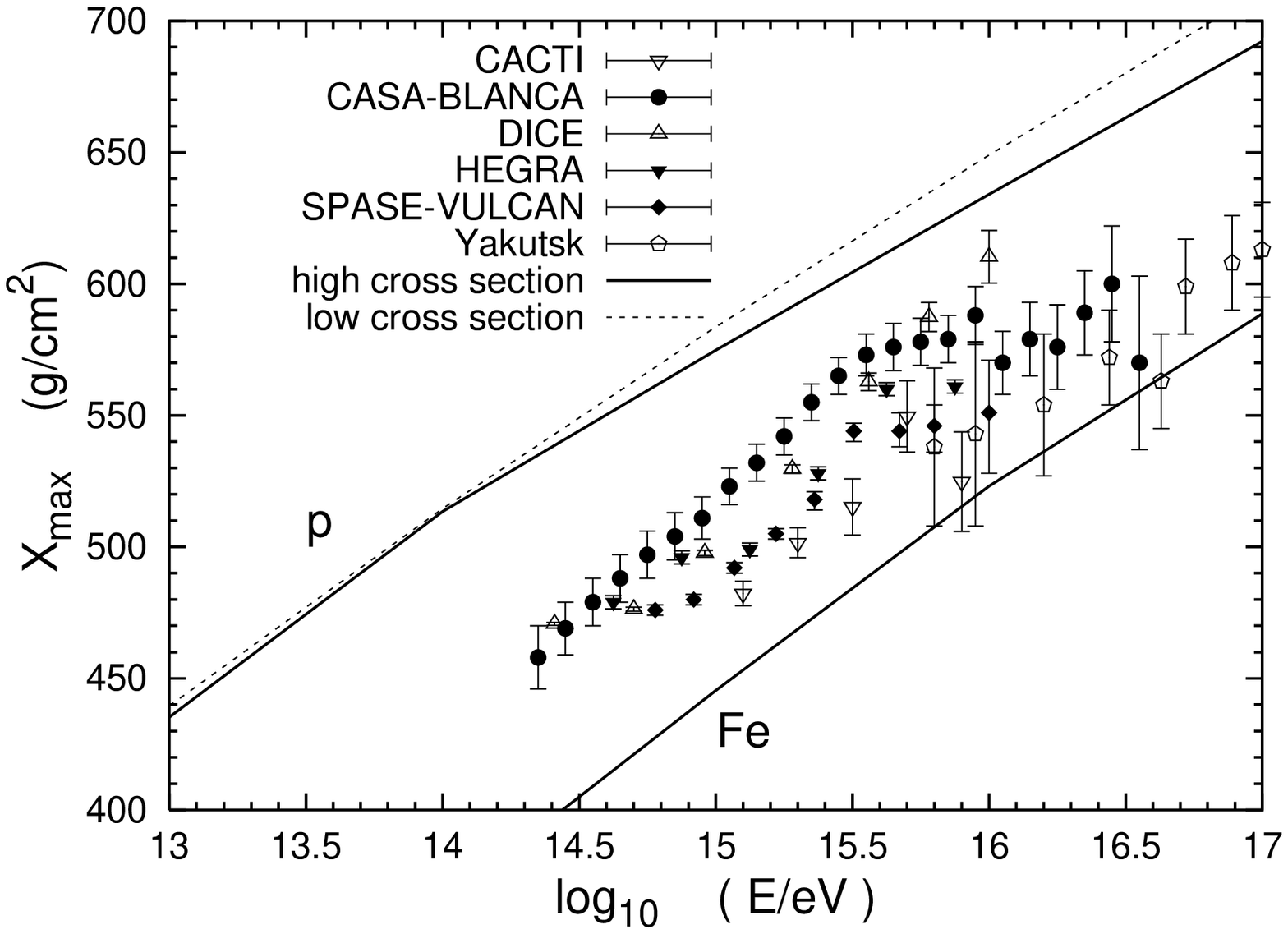}}
\vspace*{-0.8cm}
\caption{
Upper panel: two cross-section extrapolations which are compatible with
the currently available Tevatron cross-section measurements. Lower
panel: predictions for the position of the shower maximum, 
\protect$\langle X_{\rm max} \rangle$, using the two
cross-sections. For comparison also the expectation of iron-induced
showers are shown.
}
\vspace*{-5mm}
\label{fig:cs-ext}
\end{figure}
The inelastic p- and $\pi$-air cross-sections are very important
parameters of EAS simulations (see, for example, Fig.~\ref{fig:rates}). 
Unfortunately the measurements available from Tevatron allow for a wide
range of different 
extrapolations.\footnote{The CDF
Collaboration obtained with $\sigma_{\rm tot} = 81.83 \pm 2.29$ mb
a value
which is considerably greater than those reported by E710 
and E811 \cite{Tevatron-cs} ($72.81 \pm 3.1$ mb and $71.71 \pm 2.02$ mb,
respectively).} 
The arising uncertainty translates directly to predictions for air
showers.
As shown in Fig.~\ref{fig:cs-ext}, the difference in the 
$\langle X_{\rm max}\rangle$ predictions
increases to more than 20 g/cm$^2$ at $10^{17}$ eV. This difference has
to be considered as a lower limit since it corresponds only to the 
change of the proton  and pion interaction lengths 
with air and not any additional 
model changes.

An accurate measurement of the p-p cross-section at LHC by the TOTEM
Collab. \cite{Matthiae01a} would
restrict the extrapolations and hence improve the predictive power of
currently used models. An interesting option is the combination of the
TOTEM and CMS detector readout which would allow for the combined
analysis of events. In such a scenario the analysis of
leading particles would be possible, within a restricted phase space range,
in minimum bias measurements.

All LHC experiments will have the potential to contribute to minimum bias
measurements in the central region \cite{LHC-exp}. 
The phase space regions covered by 
the ATLAS and CMS detectors are very similar, $|\eta| < 2.5$ with
particle tracking and $|\eta| <5$ with hadronic and em. calorimeters.
LHCb will offer particle identification in the range $1.9 < \eta < 4.9$.
In general there is a lack of forward detectors 
(the FELIX proposal of a dedicated detector for forward measurements 
was not approved \cite{FELIX}).

One of the general problems will be the high
luminosity, causing numerous independent p-p interactions per 
bunch crossing. 
Therefore the minimum bias measurements needed for event generator
tuning have to be done when the collider starts operating and the
luminosity is still low. From the point of view of air shower physics
the acceleration of light ions would be of greatest interest, in
particular asymmetric beam configurations such as p-C. 

An example of a dedicated forward detector is the Castor project,
planned as subdetector at CMS.
It is designed to measure the ratio of
electromagnetic to hadronic energy in the forward region (app. $5.5 <
\eta < 7.2$) \cite{Angelis01a}. Options to increase
the angular and momentum resolution of this detector to enhance its
physics potential are currently studied.

There are a number of important low-energy experiments which help fill
in gaps in measured data (see, for example, \cite{Engel00a})
or improve the precision of available data. 
Whereas measurements of p-N collisions are most important 
for understanding inclusive
neutrino and muon production, pion initiated reactions dominate in EAS.

The HARP experiment, 
motivated by the physics of atmospheric neutrinos, is designed to
measure secondary hadrons, including particle identification, with virtually
full phase space coverage \cite{Barr01a}. Various particles (p,
$\pi^\pm$, and K$^\pm$) are scattered off nuclear targets including
nitrogen and oxygen. The beam energies range from 2 to 15 GeV.
A related experiment \cite{P322} took data at 100 and 158 GeV using a
modified setup of NA49.

Another dedicated low-energy experiment with full particle
identification is MIPP (E907) at Fermilab \cite{MIPP}. It
will use the main injector and allow the investigation of interactions
induced by p, $\pi^\pm$, K$^\pm$ and $\rm \bar p$ on a variety of 
nuclear targets ranging from H$_2$ to Pb. The beam energy will be
between 5 and 120 GeV. 

Some of the experiments at the Brookhaven AGS measure leading particle
distributions in p-A interactions \cite{Cole-QM2001-review}. 
These measurements were motivated by the
anomalously strong stopping power of baryons observed in heavy ion
collisions. For example, new data by the E941 Collab. on 
leading proton and neutron spectra in p-Be
collisions at 12 and 19 GeV cover almost the entire large $x$ region
\cite{Barish01a}. 

Similar studies were made at the SPS beam by NA49.
The NA49 analyzed the leading proton distribution in 
p-p, p-Pb, and Pb-Pb collisions at $158\,$GeV per nucleon by comparing 
them to $\pi$-p interactions at the same energy \cite{Rybicki02a}. 
(Unfortunately there are
no NA49 results for light target nuclei such as Be or C available so
far.)

The HERA-B experiment, designed to study B-meson physics, can also 
measure minimum bias particle production in p-C collisions. 
With its beam energy of
$920\,$GeV and particle identification in the range 
$|x_{\rm F}| \lapproxeq 0.4$ it can provide important
constraints for EAS and muon flux simulations.

Finally it should be mentioned that the precise measurement of the
inclusive atmospheric muon flux by L3+Cosmics \cite{Unger02a} gives new 
insights into the description
of forward particle production in p-air collisions in the 1 to 10 TeV
energy range. In fact, no contemporary EAS Monte Carlo model can
reproduce this measurement \cite{Engel-ISVHECRI}.


\section{Conclusions and outlook}

The current lack of methods for calculating QCD predictions for soft
particle production at high energy necessitates the use of phenomenological
models and assumptions. In modern CR experiments the uncertainty in
the simulation of hadronic interactions has become the dominant source
of systematic errors, which are difficult to estimate.
Two types of data
can help tuning interaction models, namely measurements of\\
(i) general properties of (forward) hadron production, and\\
(ii) particular predictions of models to test the underlying
assumptions.\\
Whereas the former one serves mainly the adjustment of
parameters of the models, the latter one validates the model concepts
and increases the confidence in the extrapolations.

It is clear that the measurement of minimum bias hadron 
production in proton and pion induced collisions with light nuclei tops
the priority list (i) of needs for EAS simulations. 
In particular data of
fast secondary particles, both charged and neutral, would be important.
At high energy proton or ion induced reactions are of primary
interest whereas at low energy beams of pions and kaons are better
suited. 

All Monte Carlo codes use parametrizations for the production of 
the leading baryons which are assumed to scale with energy, up to
effects due to energy-momentum conservation. These parametrizations are
tuned to p-p data at fixed target and HERA energies
and should be considered as educated guess only.
On one hand models for hadronic
multiparticle production cannot be used at low energy as the
underlying assumptions are not applicable in this range. Thus, for tuning the
high-energy extrapolation of models, p-nucleus data
at energies below 200 GeV is of limited use only. On the other hand
high-energy data of leading baryons in p-nucleus interactions
is not available.

Transverse momentum spectra of particles are mainly of interest at
low interaction energies. At high energy the scattering angle of
most of the secondary particles is negligible and does not contribute
to the lateral extent of EAS. Therefore data will be well-suited if
high-energy measurements are integrated over $p_\perp$.

The measurement of proton- and pion-nucleus cross-sections is important
as these cross-sections influence the absorption in the
atmosphere. The corresponding proton measurements at accelerators cover
the energy range up to 400 GeV but pion beam data is virtually absent.

Finally, recalling that ions in the range from He to Fe are dominating
the cosmic ray spectrum, it should be emphasized that the
measurements outlined above done with ion beams are of great interest, too.
Ideally, the measurements should integrate over all impact
parameters and not be restricted to central collisions.

The list (ii) is more model specific but the inclusive minijet cross-section 
and its energy dependence are key observables for all models.
Similarly, parton densities at low $x$ and investigations of the range of
applicability of perturbative QCD will contribute to the reliability of
the model extrapolations.

The discussions at the NEEDS workshop clearly showed that both the HEP
and the CR communities are interested in a closer
collaboration. 
However, dedicated measurements of data relevant to EAS simulation
will only be done if the data
potentially help to analyze the cosmic ray measurements with 
significantly better accuracy.
More work has to be done to make the interests and needs of cosmic
ray physics more transparent to the HEP community.

Accelerator experiments principally offer
access to their data and resources to cosmic ray colleagues. However, as is
the case in general, also projects of measurements at 
colliders are subject to evaluation, approval or rejection by
committees and therefore have to be based on well-defined physics
objectives and competitive, cost-effective designs.
Therefore any form of active involvement of institutes 
or members of the cosmic ray 
community in form of 
\begin{itemize}
\item
sending people to accelerator experiments to
help performing measurements and data analyses and
\item
financial support and cooperation
\end{itemize}
will be highly appreciated and is the best way to improve
recognition of the EAS simulation problems.

Systematic studies of the uncertainties in contemporary hadronic
interaction models are needed to work out the most sensitive observables
whose measurement will allow to improve the physics descriptions and
reduce the uncertainties of the extrapolations in energy and phase
space. The publication of the results of such investigations is
important for future reference and justifying funding proposals.


\vspace*{3mm}
\noindent
{\bf Acknowledgments:} 
The author acknowledges lively and fruitful discussions 
with the attendants of the NEEDS workshop in Karlsruhe.
He thanks in particular T.~Gaisser,
A.~Haungs, D.~Heck, S.~Ostapchenko, J.~Ranft, 
H.~Rebel, S.~Roesler, T.~Stanev, and T.~Thouw
for the collaboration on various subjects related to this work, and L.
Jones for comments on the manuscript.
The author was supported by the US Department of Energy contract
DE-FG02 91ER 40626.



\end{document}